# Microemulsion nanocomposites: phase diagram, rheology and structure using a combined small angle neutron scattering and reverse Monte Carlo approach

Nicolas Puech[1,2], Serge Mora[1], Ty Phou[1], Grégoire Porte[1], Jacques Jestin[3], Julian Oberdisse[1]*

[1] Laboratoire des Colloïdes, Verres, et Nanomatériaux (LCVN), Université Montpellier II, UMR CNRS 5587, 34095 Montpellier, France

[2] Centre de Recherche Paul-Pascal, UPR CNRS 8641, Avenue A. Schweitzer, 33600 Pessac, France

[3] Laboratoire Léon Brillouin (LLB), CEA Saclay, UMR CEA/CNRS 12, 91191 Gif-sur-Yvette, France

revised manuscript, 22/07/10

**Abstract:**

The effect of silica nanoparticles on transient microemulsion networks made of microemulsion droplets and telechelic copolymer molecules in water is studied, as a function of droplet size and concentration, amount of copolymer, and nanoparticle volume fraction. The phase diagram is found to be affected, and in particular the percolation threshold characterized by rheology is shifted upon addition of nanoparticles, suggesting participation of the particles in the network. This leads to a peculiar reinforcement behaviour of such microemulsion nanocomposites, the silica influencing both the modulus and the relaxation time. The reinforcement is modelled based on nanoparticles connected to the network via droplet adsorption. Contrast-variation Small Angle Neutron Scattering coupled to a reverse Monte Carlo approach is used to analyse the microstructure. The rather surprising intensity curves are shown to be in good agreement with the adsorption of droplets on the nanoparticle surface.

**Figures:** 11

**Tables:** 1

* author for correspondence: oberdisse@lcvn.univ-montp2.fr

## 1. Introduction

The dispersion of nanoparticles in *bulk* polymer matrices is an important aspect of all reinforcement applications [1]. Depending on the scope, individual dispersion with the highest possible interfacial area for maximal matrix-particle interaction is sought, or on the contrary the formation of a stiff particle network [2-5]. The reinforcement of *soft* polymer matrices with nanoparticles, like polymer networks in solution [6-9], raises interesting physical questions concerning the dispersion of the nanoparticles [10], the possibilities of connecting them to the polymer network [11-13], and how this influences the network structure and ultimately the network rheology [14]. The present work aims at a better understanding of the phenomena related to the reinforcement effect.

Microemulsion gels are conceptually simple and versatile realizations of aqueous polymer networks [15-17]. They are made of microemulsion droplets connected by telechelic polymer molecules, i.e. triblock copolymers having two hydrophobic sticker blocks at the extremities, which may also self-assemble in solution [18-22]. The droplets thus form the nodes of a polymer network, with usually Maxwellian rheological properties [23]. The structure can be easily studied using small angle scattering techniques, as the rather massive nodes dominate the scattering [24, 25]. Of particular interest for the present study is the possibility to incorporate bare, hydrophilic silica nanoparticles, which we have shown to be individually dispersable in microemulsion gels [26]. For such a multi-component system, Small Angle Neutron Scattering (SANS) is a powerful technique, as contrast variation can be employed to study either the matrix, or the filler, or the interaction between both [27].

In this paper, the following key-questions will be addressed: (a) What is the effect on the phase diagram of a microemulsion gel of adding well-dispersed silica nanoparticles ? (b) How are the rheological characteristics modified? Is there a specific reinforcement effect of such aqueous gels? (c) How is the structure of the microemulsion gel modified by the nanoparticles?

The outline of this article is as follows. After the experimental section (2), changes in the phase diagram of the microemulsion gel upon addition of well-dispersed silica nanoparticles are discussed in section 3.1. Throughout the article, two distinct gels, made of two different microemulsion droplets ($R \approx 30$ Å and $100$ Å), are investigated in parallel, in order to study the influence of the nanoparticle-to-droplet size ratio. The observed shifts in the phase diagrams are related to those of the rheological properties, and in particular to the shift of the percolation line, presented in section 3.2. It will be shown that the silica nanoparticles induce an increase in the Maxwell modulus G, and a decrease of the relaxation time τ. The apparition of a second, longer relaxation time related to the silica particles is also reported in this section. The structure of the silica-loaded microemulsion gels has been studied by contrast-variation Small Angle Neutron Scattering (SANS), and the results are presented in section 3.3. In the discussion of the structure, section 4, the analysis of the rather complex intensity curves has been backed up by reverse Monte Carlo simulations. Details of the latter are outlined in the appendix. In section 5, finally, simple models for the reinforcement on these microemulsion nanocomposites are developed, before concluding in the last section.

## 2. Experimental section

### 2.1 Preparation of the microemulsion gels with silica nanoparticles

Synthesis of the triblock copolymer $CH_3$-$(CH_2)_{17}$-NH-CO-O-$(CH_2$-$CH_2$-$O)_n$-CO-NH-$(CH_2)_{17}$-$CH_3$ has been described elsewhere [19, 28]. Here, we use a 10k-poly(ethylene oxide) chain (220 EO units) which has a hydrophobic $C_{18}$-sticker at each extremity, connected through a urethane group. The radius of gyration of the PEO-block is about 35 Å. Bindzil silica (B30/220, 30% wt, radius 100 Å, log-normal polydispersity ≈ 20%) suspended in water was a gift from Akzo Nobel. The pH of the delivered stock solutions was set between 9 and 10 in order to ensure maximal colloidal stability to the silica by electrostatic repulsion. All silica-containing samples have thus been prepared at pH 10, all others at pH 7. The order of mixing of the components is of importance due to the high viscosity of the microemulsion networks. First, microemulsions were formed with a surfactant (Triton X-100), a cosurfactant (octanol), and oil (octane), in water. Then the colloidal silica solution was added, and finally the telechelic polymer. The variable characterizing the connectivity is the amount of copolymer, due to the obvious bridging properties of the telechelic polymer. It is defined as the average number of stickers per droplet (e.g., one polymer per droplet gives r = 2). Two different microemulsions gels have been prepared: system A made of small swollen micelles, and system B made of bigger microemulsion droplets.

## 2.2 Rheology

Storage G' and loss moduli G'' were determined using a strain-controlled rheometer ARES-RFS (TA Instruments) with cone-plate geometry (radius 2 cm, cone angle 0.02°). Oscillatory shear experiments have been performed as a function of angular frequency $\omega$ in the linear regime (low amplitudes, $\gamma = 10\%$), at T = 22°C, and we have checked linearity up to $\gamma = 20\%$. The reinforcement factor presentation was chosen to highlight the filler contribution to the stress. This presentation consists in normalising the filled gel modulus $G_f$ by the one of the pure matrix $G_m$ under otherwise identical conditions. This quantity thus gives the increase in modulus due to the filler particles.

## 2.3 Small Angle Neutron Scattering

Scattering experiments have been performed on beamline PACE at LLB (Saclay). The configurations used were: 1m and 4.68 m at 6Å; 4.68 m at 13Å. Samples have been prepared in $D_2O$, small quantities of $H_2O$ from the silica stock solution have been accounted for in the calculation of the solvent scattering length densities. Empty cell scattering has been subtracted, and detector efficiency has been corrected with 1mm-light water scattering. Data were normalized to obtain absolute units $(cm^{-1})$ by an independent measurement of the incoming flux. The incoherent background subtraction was performed by comparing to Porod scattering expected for smooth interfaces. Note that our gels are special networks where the polymer molecules at the origin of the network elasticity are not easily visible by scattering techniques, like it would be the case for pure polymer networks. Due to the difference in scattering between the tenuous polymer molecules and the dense microemulsion droplets making up the nodes of the network, only the nodes are visible in a straightforward scattering experiment. Using independent contrast variation experiments, we have determined the scattering length density of silica $(3.5 \ 10^{10} cm^{-2})$, of the swollen micelles $(0.2 \ 10^{10} cm^{-2})$, and of the microemulsion droplets $(-0.3 \ 10^{10} cm^{-2})$. In the silica-filled gels, nodes, silica nanoparticles, or both can thus be probed by adjusting the solvent scattering length density using $H_2O/D_2O$ mixtures.

### 3. Results

### 3.1 Phase diagrams

The experimental phase diagram of our microemulsion system depends on five concentrations, which can be rationalized in three independent groups. The shape and size of the microemulsion droplet is governed by the cosurfactant-to-surfactant ratio $\Omega$, which sets the curvature of the hydrophilic-hydrophobic interface, and the oil-to-surfactant ratio $\Gamma$. The latter defines the degree of filling of the droplets with oil. Note that these two parameters can not be varied independently for droplets [26]. We have chosen to study two microemulsions of fixed droplet size, one made of small, hardly swollen micelles (system A, $\Omega = 0$, $\Gamma = 0.05$, R = 32 Å), and the other one made of larger microemulsion droplets (system B, $\Omega = 0.4$, $\Gamma = 0.7$, R = 104 Å). The study of two systems probes the influence of the ratio between droplet size and radius of gyration of the macromolecule, as suggested by recent simulations [29]. Once the droplet size is chosen, we are left with three parameters: the volume fraction of droplets $\Phi_m$, the one of silica nanoparticles $\Phi_{si}$, and the amount of telechelic copolymer given by the connectivity r. In Figs.1 and 2, two cuts through the phase diagram, one at $\Phi_{si} = 0$, the other at $\Phi_{si} = 6.54$ %, are superimposed, as a function of the two remaining parameters, $\Phi_m$ and r. In the absence of silica nanoparticles ($\Phi_{si} = 0$), the usual phase diagram as discussed in ref. [17] is found, with three regions: at low $\Phi_m$ and low r, the droplets are decorated by the copolymer, the so-called flower conformation. As one increases $\Phi_m$, the fluid samples - with a viscosity essentially governed by the diffusion of droplets (or possibly some aggregates) - develops suddenly into a highly viscous, elastic gel. The viscosity of this gel would be infinite if the links introduced by the polymer were permanent. However, due to the finite disengagement time of the polymer stickers out of the micelles, the system remains fluid, albeit highly viscous. In analogy with percolating links or nodes on lattices, this transition is usually referred to as percolation, and it is represented in Fig. 1 and 2 by the dotted and broken lines. The third region of interest is the two-phase region found at high r. It results from the strong attractive interactions at high polymer content, leading to a dense gel-phase coexisting with the dilute flower-phase.

When comparing Figs. 1 and 2 in absence of silica, it appears that the two systems behave in a very similar manner. The location of the two-phase region and the percolation line are shifted in droplet concentration, the smaller droplets percolating at lower volume fractions. This is consistent with the fixed molecular mass of the copolymer, which sets the range of the attractive interaction, roughly given by the radius of gyration of the PEO [29-32]. Droplets are correlated with a typical surface-to-surface distance $R_g$, which sets the typical volume fraction of well-percolated systems to 4% for swollen micelles (A), and to 20% for the bigger droplets (B), respectively.

Upon addition of silica nanoparticles, the two-phase regions in the diagrams are shifted towards lower concentrations, and the percolation line follows the same trend, accompanied by a shift to lower connectivity. This implies, e.g., that a low-viscosity sample located close to the percolation line gels as silica is added. The *hydrophilic* and well-dispersed silica beads thus participate in the network formation, which is itself based on the *hydrophobic* interaction between stickers and droplets. In order to understand this intriguing phenomenon, we have undertaken rheological studies along the two arrows plotted in the phase diagrams.

### 3.2 Rheological properties of silica-loaded microemulsion gels

The striking rheological properties of pure and silica-containing gels have been characterized using oscillatory shear experiments in the linear domain. The resulting moduli, G'($\omega$) and G''($\omega$), have been found to be essentially Maxwellian [26], i.e. determined by a single shear modulus G and a single relaxation time $\tau$. Deviations from such ideal behaviour will be discussed later. Several sample series have been measured, either along a line of fixed droplet volume fraction $\Phi_m$, varying the connectivity r, or along a line at fixed r, varying $\Phi_m$. In both cases, the samples evolve from a phase of low viscosity to a highly viscoelastic phase, the transition taking place at the percolation threshold. The evolution with r has been investigated recently [26], and both G and $\tau$ have been found to be nicely described by a percolation approach:

$$F(r) = F_o \left( \frac{r - r_c}{r_c} \right)^{\beta} \tag{1}$$

where F represents either G or $\tau$, $r_c$ is the percolation threshold, and $\beta$ the critical exponent. The fit parameters describing their evolution *with connectivity* are summarized in Table 1 for the two systems under study. It appears that in both systems the addition of silica nanoparticles decreases the percolation threshold by more than a factor of two, i.e. there is a strong effect of the nanoparticles. These results underline that the modulus is reinforced, and that the relaxation time is decreased by the addition of nanoparticles, for both systems under study. One may note that the values of the percolation threshold $r_c$ characterize the onset of percolation, and that they represent average numbers of stickers per droplet. Our findings of rather low values indicate that percolation is a spatially heterogeneous process, as obviously more than one molecule per nanoparticle is needed to connect all droplets other than in a row. Similar behaviour has been observed in numerical simulations of such gels [33]. Moreover, the presence of silica clearly favours the effective connectivity, in the sense that is increases the probability of bridging. Possible reasons for this effect will be discussed in the light of the neutron scattering experiments.

| System | Without silica | | | With silica (6.54%v) | | |
|---|---|---|---|---|---|---|
| | $G_0, \tau_0$ | $r_c$ | $\beta$ | $G_0, \tau_0$ | $r_c$ | $\beta$ |
| **(A) R= 32 Å** **Modulus** | 250 Pa | 2.3 | 1.6 | 200 Pa | 1.0 | 1.6 |
| **(A) Relaxation time** | 8.5 ms | 2.5 | 1.6 | 1 ms | 1.0 | 1.6 |
| **(B) R= 104 Å** **Modulus** | 120 Pa | 3.5 | 1.6 | 32 Pa | 1.5 | 1.6 |
| **(B) Relaxation time** | 6.7 ms | 3.5 | 0.65 | 4 ms | 1.5 | 1.6 |

*Table 1: Fit parameters of percolation eq. (1) as a function of connectivity r, at fixed droplet volume fraction.*

The location of the percolation lines displayed in the phase diagrams has been obtained from the rheological measurements. In the literature, percolation is usually discussed in terms of the *concentration of percolating objects* or nodes [34], whereas we have up to now (Table 1) investigated a system where the density of nodes was constant and the connectivity increased by adding polymer. For the purposes of the present paper, we have performed several series of experiments at fixed connectivity r, and varied the droplet volume fraction $\Phi_m$. As before, the modulus G has been deduced from G' and G", and the results are plotted in Figure 3 for system A, and in Figure 4 for system B. The modulus is found to increase as the silica is added, and it still follows a percolation law, with the volume fraction $\Phi_m$ replacing the connectivity r in eq.(1). From the fit parameters given in the captions, it is obvious that the percolation threshold $\Phi_c$ is shifted to lower volume fractions, namely from 1.0% to 0.8% in the case of the small micelles, and from 7.5% to 6% in the case of the bigger microemulsion droplets. Note that equivalent fits can be obtained with slightly different parameters, the error bars being of the order of 0.1% (resp. 1%) for the percolation threshold of system A (resp. B), and about 0.1 for the exponent. We have already pointed out that the order of magnitude of these threshold concentrations of pure microemulsions can be estimated from the radius of the droplets and the $R_g$ of the PEO, supposing that bridged particles must typically be at a surface-to-surface distance $R_g$ [29-32, 35]. Its shift with added nanoparticles, however, escapes from such simple considerations. On the other hand, the exponent is not affected by the presence of silica, and its value changes only slightly from 1.9 to 1.8 when increasing the droplet size. Interestingly, the value of the exponent is compatible with the one predicted by Surve et al for networks made by polymer bridging (1.8), which they argue to be general for random percolation models and entropic gels [34].

In the context of mechanical reinforcement of polymer networks by nanoparticles, the important question is not so much the absolute value of the modulus, but by how much it is increased upon addition of nanoparticles. Although there is currently no general theory available for particulate reinforcement, this concept is obvious from existing limiting laws for reinforcement by spheres at low concentration [36-38]:

$$G_f = G_m \ [\ 1 + 2.5\ \Phi_{si} + O(\Phi_{si}{}^2)\ ] \qquad (2)$$

where the Maxwell modulus of filled networks is denoted $G_f = G(\Phi_{si}=6.54\%)$ and the one of the unfilled matrix $G_m = G(\Phi_{si}=0)$. This equation offers a different point of view on both the shift of the percolation threshold and the increase of the modulus. In Figure 5, we have plotted the ratio $G_f/G_m$ as a function of droplet volume fraction $\Phi_m$, for both systems under study. At high droplet concentrations, this reinforcement factor displays a plateau value which reflects the reinforcement ability of the dispersed nanoparticles. The reinforcement here is larger than the purely 'Einsteinian' value ($G_f/G_m = 1.16$), which indicates that the phenomenon can not be of hydrodynamic origin only. At the approach of the critical droplet volume fraction $\Phi_c$, the reinforcement factor increases considerably: for the swollen micelles, the upturn is located at $\Phi_c = 1.0\%$, for the bigger ones at 7.5%. Technically, this upturn is caused by the vanishing denominator $G_m$, which decreases more rapidly than the numerator $G_f$. Physically, it means that close to the percolation threshold, the contribution of the silica particles to the network formation becomes dominant, and it makes the system cross the percolation line. Figure 5 thus illustrates not only the 'normal' reinforcement of systems far from percolation, but also the nanoparticle effect close to the percolation. This illustrates again that the silica nanoparticles contribute to the network formation.

Up to now, we have focussed on the evolution of the *modulus G* of the filled microemulsion gels as determined by the Maxwell fit of G' and G''. Now, we turn to the effect of silica on the second parameter of the Maxwell fit, the *relaxation time* $\tau$. We have regrouped in Figure 6 the relaxation time as a function of $\Phi_{si}$ for the two systems under study here, at different connectivities r. One may note that the influence of r is to shift the relaxation time, more polymer molecules leading to a slower relaxation, whereas the general dependency of $\tau$ on $\Phi_{si}$ seems unaffected. Surprisingly, the relaxation time is seen to decrease systematically with $\Phi_{si}$, and a linear fit as given in the plot shows that the decrease is proportional to $(1-3.6\ \Phi_{si})$ and $(1-\Phi_{si})$ for systems A and B, respectively. The effect of silica on the relaxation time is thus considerably stronger for system A than for the bigger droplets of system B.

The behaviour of the pure microemulsion gels is almost perfectly Maxwellian. Upon dispersion of silica nanoparticles, however, deviations occur, as already noted in our previous articles [26, 39]. In Figure 7, a fit of G' and G'' with a double Maxwell model, with two moduli and two relaxation times, is shown to yield a satisfying result for the microemulsion gel (system A, $\Phi_m$ =0.04, r=10, $\Phi_{si}$ =5.7%v), and the same has been observed for system B as shown in the ESI. The value of the principal relaxation time $\tau_1$ is rather small as before (0.05 s for this sample), whereas $\tau_2$ is longer, of the order of 0.2 s. Note that the exact value of $\tau_2$ is subject to caution, as slightly different values yield essentially identical fits. The modulus corresponding to the second mode, $G_2$, is much weaker (170 Pa) than the principal modulus, $G_1$ (4600 Pa). In the inset of Figure 7, we have plotted the dependence of $G_2/G_1$ on the silica volume fraction. This ratio, which reflects the relative importance of the second relaxation mode, is seen to increase with $\Phi_{si}$. To fix ideas, a linear fit of the complete data set yields a prefactor of 0.5 for the swollen micelles, and 1.7 for the microemulsion gel (not shown). This proportionality suggests that the second relaxation is related to the silica nanoparticles.

To summarize this discussion on relaxation, we have observed that the principal relaxation time decreases with increasing nanoparticle content in both microemulsion systems. In parallel, a second, weaker and much slower relaxation mode shows up with increasing silica content.

### 3.3 Structure of silica-loaded microemulsion gels

The dispersion of *silica* nanoparticles in the microemulsion gel phase has been studied previously by SANS using contrast variation, i.e. in a solvent mixture matching the organic phase [26]. For both the swollen micelles and the bigger droplets, the silica dispersion was found to be identical to the one of a pure silica suspension in the absence of copolymer (r=0), and only slightly modified in the case of networks (r=10). Note that the pure silica suspension is perfectly dispersed due to electrostatic repulsion, which leads to a peak around 0.017 $\text{Å}^{-1}$ at $\Phi_{si}$ =6.54%: the nanoparticles are thus not destabilized by the microemulsion gel.

In this article, the structure of the nanoparticle-containing *microemulsion gels* has been studied in a similar manner, by using a solvent mixture index-matching the silica nanoparticles. Then only the micelles or droplets contribute to the scattering. In Figures 8 and 9, the structures of system A and B with matched particles as seen by SANS are shown.

In Figure 8a, the volume fraction of silica particles is increased from 1% to 6%v in a gel of swollen micelles (system A, r=10). The intensity is seen to be unchanged at large angles, which means the surface covered by the surfactants stays the same. It also implies that the silica is well matched, as its surface would otherwise contribute to the scattering. The micellar interaction peak located around 0.05 $\text{Å}^{-1}$ remains constant, indicating that the micelles stay at the same distance from each other. At lower angles, however, a second peak shows up, and in the linear representation displayed in the inset it is seen to be rather dominant. Surprisingly, the height of the peak appears to be correlated with the silica volume fraction, and its position close to where the silica structure factor peak would be (0.02 $\text{Å}^{-1}$), if the silica were visible.

In Figure 8b, we have superimposed the scattering of the pure systems to one of the intensities of Figure 8a (6% silica). The pure swollen micelles rescaled to identical contrast are seen to compare nicely with the micelles in the presence of silica, but do not possess the additional maximum at small angles. We have also superimposed the intensity of a pure silica solution (6%, shifted in I by a factor of 0.10), and it is indeed seen to have the maximum at the same position as the contrast-matched data. This suggests that the silica restructures the microemulsion gel, and thereby becomes visible although it has no scattering contrast.

The analogous situation for the bigger droplets of system B is shown in Figure 9. Again, the volume fraction of silica particles is increased (from 1% to 5%v) in the gel phase (connectivity r=10). The intensity remains unchanged at large angles, but the interaction peak around 0.025 $\text{Å}^{-1}$ now changes: it increases, as it is also displayed in the inset of Figure 9a. At lower angles, no additional feature like a second peak shows up. In Figure 9b, we have again superimposed the pure systems to one of the intensities of Figure 9a (5% silica). The pure swollen micelles rescaled to identical contrast are seen to compare nicely with the micelles in the presence of silica. We have also superimposed the intensity of a pure silica solution (5%) in $D_2O$. This time, the maximum of the pure silica is not visible in the intensity of the gel. As a last comparison, the pure gel (no silica) has also been rescaled to identical contrast conditions and plotted. Its peak is far more to the right. It thus looks like the effect of the polymer is to establish a closer contact between the droplets, inducing a peak-shift to the right, whereas the effect of the silica is in the opposite direction. The result is a structure which closely resembles the one in absence of both silica and polymer, as shown by the nice superposition with this data set. It thus appears that the effect of the silica is to perturb the structure of the gel, inducing on average larger inter-droplet distances.

To summarize, a reorganization of the gel due to the matched silica is observed in both systems. In the case of the bigger droplets there is a shift to smaller q-vector and enhancement of the interdroplet peak around 0.025 $\text{Å}^{-1}$, whereas the silica structure is imprinted onto the gel of small micelles.

## 4. Modelling the structure

We have undertaken numerical simulations in order to investigate the curious double-peak structure shown in Figure 8. As we have checked by independent contrast variation (cf. ESI), the silica is well-matched; nonetheless, there is a peak located at the silica peak position. Two simple explanations are possible. Either the presence of the silica nanoparticles in the microemulsion gel is sufficient to generate a correlation hole in the micellar structure factor, by excluding the micelles from the volume occupied by the silica, or the micelles rearrange around the silica such that they generate a density profile highlighting the silica. This is conceptually very close to the observation by scattering of pore structure of index-matched material by adsorbed surfactants [40]. We have checked both scenarios, and the results are shown in Figure 10.

The first prediction has been generated using a Reverse Monte Carlo (RMC) algorithm as outlined in the appendix. We have fitted the scattering of a gel of swollen micelles in absence of silica, using 8000 beads in a simulation box. The result is a configuration of bead positions the scattering of which exactly matches the pure gel case. Secondly, we have run an independent RMC-simulation for the pure silica, and obtained a configuration of nanoparticles satisfying the silica scattering. In a last, unifying step, we have copied the silica particles into the micellar simulation box, and suppressed all micelles which where located at positions occupied by silica. Then the scattering function of the remaining micelles was recalculated, and the result is shown in Figure 10. As one can see, the effect of the presence of silica on the scattered intensity is negligible, and the small-angle peak is not produced by this correlation-hole scattering. Intuitively, this can be understood from the rather high dilutions: decreasing the amount of available space (96% for a 4% micellar gel) by 6% due to the presence of silica does not affect the correlations.

The second prediction is more elaborate to verify. We have adapted our RMC-code to describe the situation, using the previously determined silica structure in agreement with the pure silica scattering. Then, micelles were placed – respecting the excluded volume – randomly in the simulation box already filled with silica. Again, RMC was used to fit the scattering data of Figure 8 (6%v), now with the constraint of the presence of invisible silica in the simulation box. The resulting fit is also shown in Figure 10, and the agreement with the data is very satisfying. In itself, this result means that swollen micelles can be arranged in such a way around suspended silica particles that they reflect the imprint of the nanoparticles in scattering. If one wishes to know more about this particular micellar structure, one has to analyze the pair correlation functions, and in particular the cross-correlation term between silica and micelles. As a result, it is found that there is a significant local increase of micellar density close to the silica surface, in the range of one to two micellar radii, by 12%. This strongly suggests that the silica particles are decorated by a layer of micelles, which can then serve as an anchoring point to the hydrophobic polymer stickers.

In this context, it may be interesting to point out that we had previous characterized the adsorption of non ionic surfactant molecules on silica in solution [41-43]. In particular, the adsorption of micelles of TX-100, the surfactant used for the microemulsion formulation of the present study, on identical silica beads has been determined in low-concentration studies [41, 44]. About 15 micelles were found to be adsorbed to each silica nanoparticle. These measurements give additional credit to the hypothesis of adsorption of microemulsion droplets on the silica particles, thereby connecting them to the network of the microemulsion gel.

## 5. Modelling the reinforcement factor

One can use the reinforcement factor of the modulus of microemulsions well within the gel region to estimate the binding strength of the micelles to the silica. It is well known that the elastic modulus of homogeneous gels increases with the connectivity r ($\Phi_m$ being fixed). Far from percolation, simple network theory allows the estimation of the modulus G assuming a homogeneous distribution of links:

$$G = \nu \, kT \qquad (3)$$

where $\nu$ is the number density of active links, k is the Boltzmann constant and T the temperature. In reality, only a fraction of copolymer molecules are active links [26, 29]. Far from percolation, the reinforcement factor $G_f/G_m$ is found to become independent of r and $\Phi_m$. As shown recently, its value increases with $\Phi_{si}$ proportionally to $1+18\Phi_{si}$ for system A (resp. $1+12\Phi_{si}$ for system B) [17]. If we relate this increase to the silica particles, one may deduce the equivalent number of polymer molecules P which would replace one silica particle. For system A, we find:

$$G_f = G_m \ (1+18 \ \Phi_{si}) = G_m + P \ kT \ N_{si}/V \qquad (4)$$

where $N_{si}/V$ is the number density of silica particles. Using typical values ($G_m = 2400$ Pa) and $N_{si}/V = \Phi_{si}/V_{si}$, we find $P \approx 44$. An analogous calculation for system B yields a much lower P $\approx 4$. Given that there are fewer polymer molecules in system B (which has therefore a lower modulus), and that the observed reinforcement is also lower, the difference is not surprising. Numerically, in the case of system A, the number of 44 suggests that each of the adsorbed micelles is bound with a strength corresponding to about three polymer molecules, which appears to be a reasonable value. In the case of system B, a much lower number of adsorbed droplets, one or two, can be hypothesized. Finally, as $G_m$ scales with the droplet number density using eq.(3), one may introduce the ratio of the silica to droplet volume, when solving eq.(4) for P. This ratio varies by about a factor of 27 from system A to system B. A posteriori, this justifies the exploration of different size ratios.

Another way of rationalizing the increase in modulus is to ascribe it to the increase in the effective total connectivity due to addition of silica particles. Redistributing the 'virtual' additional new links due to the silica over all droplets leads to the following effective increase in connectivity r:

$$\Delta r = 2P \frac{\Phi_{si}}{\Phi_m} \left( \frac{R}{R_{si}} \right)^3 \qquad (5)$$

where $R_{si}$ and R are the silica bead and droplet radii, respectively. For system A ($\Phi_{Si}$ =6.5% and $\Phi_m$ =4%), one finds $\Delta r = 3.8$, and for system B ($\Phi_{Si}$ =6.5% and $\Phi_m$ =20%), $\Delta r = 2.6$. It is interesting to note that the increase in r moves the system further from the percolation, which is in qualitative agreement with the observed shift of the percolation line to lower droplet volume fractions.

A more microscopic point of view is now proposed in order to understand the shift in percolation with silica, which is exemplified by the behaviour of the reinforcement factor as a function of droplet volume fraction $\Phi_m$ (cf. Fig. 5). We start again from the observation that some droplets are absorbed on silica particles. Telechelic polymers can bridge non-adsorbed and adsorbed droplets, and thus the elastic network can connect silica particles. Let D be the characteristic distance between silica particles, which is directly related to their number density. In presence of silica, elastic clusters of average size D are large enough to bridge two silica particles. An *infinite* cluster formed by silica particles linked by *finite* elastic clusters can thus emerge. Close to percolation, the cluster size d is expected to diverge following a classical law:

$$d = \frac{\delta}{\left[ \frac{\Phi_c - \Phi_m}{\Phi_c} \right]^\alpha} \qquad (6)$$

δ being a microscopic distance of the order a droplet radius, $\Phi_c$ the percolation threshold without silica and α an exponent of order of unity. Then, the condition on $\Phi_m$ to obtain elastic clusters sufficiently large to link silica particles becomes (see Fig. 11):

$$\Phi_m > \Phi_c \left[ 1 - \left( \frac{\delta}{D} \right)^{1/\alpha} \right] \qquad (7)$$

The clusters will bridge the silica particles as soon as d becomes comparable to D. One then obtains a new percolation threshold for the droplets volume fraction:

$$\Phi_c^{si} = \Phi_c \left[ 1 - \left( \frac{\delta}{D} \right)^{1/\alpha} \right] \qquad (8)$$

Note that the new threshold is smaller than the previous one, which is consistent with experiments (cf. Figs. 1, 2). For system A (see Fig. 3), we have measured $\Phi_c^{Si} = 0.8\%$ and $\Phi_c = 1.0\%$, leading to $D/\delta \approx 5 > 1$, which is consistent with the picture of clusters involving a large amount of droplets. For system B (Fig. 4), one measures $\Phi_c^{Si} = 6\%$ and $\Phi_c = 7.5\%$, leading to the same value $D/\delta \approx 5$. Knowing that the radius of droplets in system B and silica particles are similar, the discussion based on microscopic percolating clusters appears to be less suited for system B than for system A.

We now consider, within this last description, the elastic modulus. Due to percolation, the elastic modulus is not proportional to the polymer concentration, or, for a fixed connectivity, to the droplet volume fraction: one observes the law (without silica, cf. eq. (1) and Figs. 3, 4):

$$G_m = G_0 \left[ \frac{\Phi_m - \Phi_c}{\Phi_c} \right]^{\beta} \qquad (9)$$

We have already pointed out that similar fits than those shown in Figs. 3 and 4 can be obtained with slightly different parameters. The ones obtained show that it is possible to describe the data with the same elastic modulus and exponent. Within our picture where adsorption modifies the threshold but only slightly the network between silica, one expects the thus the same law in presence of silica, with the same elastic modulus and the same exponent, but with the shifted percolation threshold. We may then write for the filled modulus:

$$G_f = G_0 \left[ \frac{\Phi_m - \Phi_c^{si}}{\Phi_c^{si}} \right]^{\beta} \tag{10}$$

By dividing eq. (10) by eq. (9), a prediction for the reinforcement factor depending only on the shifts in percolation and the exponent $\beta$, but not on the modulus, is obtained. This prediction is plotted on Fig. 5, in good agreement for both systems A and B.

As a last point, we now consider the relaxation time behaviour. It has been shown above that the influence of silica particles can be seen to act as additional polymers. Within this description the total number of network nodes (=droplets) is constant and two types of connections between microemulsion droplets are involved: connections by polymers or by silica particles. The supplementary relaxation time (Fig. 7) that emerges in presence of silica probably comes from the later, i.e. it is a time characterizing the silica-droplet interaction. The principal relaxation time comes from elastic clusters made of polymers and droplets. It decreases with the volume fraction of silica particles. The volume fraction of droplets in these clusters is slightly smaller than in the gel without silica because some droplets are bound to the silica particles by adsorption. On the other hand, it is obvious that the viscoelastic relaxation time of the gel *without silica* increases from an unmeasurably low value at the percolation transition to its value in the gel. It is thus possible to relate the observed decrease with $\Phi_{si}$ of the principal relaxation time to the decrease of the droplet concentration in the elastic clusters. This provides us with a microscopic understanding of the influence of silica nanoparticles on the relaxation time (Fig. 6), based on droplet adsorption.

In this context, it is interesting to point out that the relaxation time of a surfactant-free system made with telechelic copolymer with a longer hydrophilic chain (PEO 35k) was found to *increase* with silica volume fraction, proportionally to $(1+\Phi_{si})$ [39]. In this case, steric hindrance due to the excluded volume of the silica may prevent the sticker from exploring other droplets, and thus increase the relaxation time. The important point is that the absence of adsorbed hydrophobic compartments on the surface of the silica, which would tend to decrease $\tau$ following our above argumentation, seems to be directly related to the absence of surfactant in this system. This suggests that the surfactant is the driving force for the formation of a hydrophobic compartment on the surface of the silica nanoparticles.

## 6. Concluding remarks

To summarize, there is strong evidence from both rheology and neutron scattering for the formation of hydrophobic compartments, presumably adsorbed micelles, on the surface of the silica nanoparticles. This explains the observed shifts in the phase diagram, where both percolation line and two-phase regions move to smaller microemulsion concentrations. It suggests that silica connected to the network via bound micelles plays the role of anchors to the droplet-copolymer clusters. The first consequence is that fewer microemulsion droplets are needed to reach percolation through finite-sized clusters. In addition, far from percolation, a reinforcement effect is obtained, caused by an increase of the density of active links. Micellar binding to silica nanoparticles is also invoked to explain the evolution of the principal relaxation time, via the decrease of the number of micellar nodes of the network.

The reinforcement has been characterized by rheology, and one may note that there is a specific effect in these originally Maxwellian systems: both the modulus and the relaxation time are affected by the silica. This situation is in contrast with mechanical reinforcement of bulk polymers, focussing on the modulus only, or the one of viscous liquids, which concentrates on the viscosity $\eta$, as exemplified by the Einstein formula (eq.(2)). In our systems, both the modulus G and the relaxation time $\tau$ evolve, thus leading to an overall change of the viscosity, because $\eta$ is given by the product of G and $\tau$ in Maxwellian systems. To finish, minor but clearly identified deviations from purely Maxwellian behaviour are caused by the incorporation of the silica in the microemulsion gel. At the current stage, the origin of the additional, longer relaxation time, which is accompanied by a small modulus roughly proportional to the silica volume fraction, is not resolved, but it is plausible that this contribution is a signature of the binding of the droplets to the silica nanoparticles.

**7. Appendix : Reverse Monte Carlo simulations**

Reverse Monte Carlo (RMC) has been applied to different types of scattering data over the past 20 years [2, 45-48]. For our purposes in small angle scattering, it reduces to an automatic fitting procedure of intensity curves by an ensemble of spheres. In *direct* modeling, a certain structure or interactions are assumed and the compatibility of its Fourier transform with the experimental data is checked [49]. In reverse modeling, no assumption on the structure or interaction potential is made. A system of spheres representing micelles or particles is rearranged randomly, and after each random step the corresponding predicted scattering is compared to the data. If the agreement is better, then the step is kept, whereas a Boltzmann criterion is used to decide on acceptance if the agreement is worse. If the simulation converges, the outcome of the simulation is a three-dimensional representation in real space of an ensemble of spheres reproducing the experimental scattering. Alternatively, one can also focus on the structure factor directly.

In practice, the following steps are executed:

- An initial configuration of N spheres is a simulation box of dimension L is produced, respecting the experimental volume fraction and if desired the excluded volume interactions. L has to be chosen such that the entire q-range is well represented. No dependence on the initial configuration has been observed (lattice structure, random,…).
- The structure factor corresponding to the current ensemble of spheres is calculated using the Debye formula. It can be used to determine the intensity, via its factorization into form and structure factor. The form factor must be known from, e.g., independent low-concentration measurements.
- The deviation between experimental and predicted scattering is determined as the $\chi^2$, i.e. the sum of the differences between the experimental and the predicted intensity or structure factors. Error bars can be included.
- The heart of the simulation is the Monte Carlo step of a randomly chosen particle. We have chosen to alternate stochastically two types of steps. The first one is simple Brownian motion, which is coupled to occasional large random jumps. This favors radical rearrangements and avoids possibly jammed situations at larger volume fractions.

- $\chi^2$ is calculated again and the last step is accepted if $\chi^2$ decreases, or increases by a small amount determined by a Boltzmann criterion with a success probability of exp(-$\chi^2$/B). This can be coupled to a simulated annealing process, which consists in a continuous decrease of B.
- By repeating the last two points a large number of times, the Monte Carlo steps are iterated.

For the simulation of the system A presented in this article, 8000 spheres of radius 32 Å at 4%v are used. This leads to a box size of about 3000 Å, and we have checked by a direct calculation that the form factor oscillations of this box are negligible in the q-range of interest. In the case of simulations in presence of silica particles, positions of these have been determined independently using a similar RMC-code applied to the scattering of pure silica solutions, taking polydispersity into account. In all cases, excluded volumes of micelles and silica nanoparticles have been respected at all times.

**Acknowledgements:** Beamtime at Laboratoire Léon Brillouin (CEA/CNRS) is gratefully acknowledged. Work conducted within the scientific program of the European Network of Excellence *Softcomp*: 'Soft Matter Composites: an approach to nanoscale functional materials', supported by the European Commission. Silica stock solutions were a gift from Akzo Nobel.

**Table caption**

**Table 1:**      Fit parameters of percolation equation (1) as a function of connectivity r, at fixed droplet volume fraction.

**Figure Captions**

**Figure 1:**      Cut through the phase diagram of swollen micelles (system A, $\Omega = 0$; $\Gamma = 0.05$) as a function of volume fraction of microemulsion $\Phi_m$ and network connectivity r. $\Phi_{si} = 0$: White two-phase region and broken percolation line. $\Phi_{si} = 6.54\%$: Shaded two-phase region and solid percolation line.

**Figure 2:**      Cut through the phase diagram of microemulsion droplets (system B, $\Omega = 0.4$; $\Gamma = 0.7$) as a function of volume fraction of microemulsion $\Phi_m$ and network connectivity r. $\Phi_{si} = 0$: White two-phase region and broken percolation line. $\Phi_{si} = 6.54\%$: Shaded two-phase region and solid percolation line.

**Figure 3:**      Modulus G (in Pa) as function of volume fraction of swollen micelles $\Phi_m$ (system A, r = 7). The results for unfilled (empty symbols, $\Phi_{si} = 0$) and filled (plain symbols, $\Phi_{si} = 6.54\%v$) micellar gels are superimposed. Fitting parameters for the modulus are $G_o = 300$ Pa, $\Phi_c = 1.0\%$, $\beta = 1.9$ (unfilled), and $G_o = 300$ Pa, $\Phi_c = 0.8\%$, $\beta = 1.9$ (filled).

**Figure 4:**      Modulus G (in Pa) as function of volume fraction of microemulsion droplets $\Phi_m$ (system B, r = 10). The results for unfilled (empty symbols, $\Phi_{si} = 0$) and filled (plain symbols, $\Phi_{si} = 6.54\%v$) microemulsion gels are superimposed. Fitting parameters for the modulus are $G_o = 150$ Pa, $\Phi_c = 7.5\%$, $\beta = 1.8$ (unfilled), and $G_o = 150$ Pa, $\Phi_c = 6.0\%$, $\beta = 1.8$ (filled).

**Figure 5:** Reinforcement factor $G_f/G_m$ of the microemulsion gels by silica nanoparticles at 6.54%v as a function of microemulsion volume fraction $\Phi_m$. The reinforcement factors of system A (r=7) and system B (r=10) are superimposed. The horizontal solid line represents the Einstein/Smallwood hydrodynamic reinforcement. The fitting curves have been obtained by combining eqs. (9) and (10).

**Figure 6:** Relaxation time $\tau$ as a function of the silica volume fraction for three microemulsion gels: system A, r=7 (◯), and system B, r=10 (▢) and 15 (△). Linear fits are also plotted and equations are given. $\Phi_m = 0.04$ and 0.20, respectively.

**Figure 7:** Storage (◦) and loss (▫) modulus as a function of $\omega$ of a microemulsion gel (system A, $\Omega = 0$; $\Gamma =0.05$, $\Phi_m = 0.04$, r = 10) containing 5.7%v of silica nanoparticles. Departures from Maxwellian behaviour are emphasized in G" at high $\omega$, and in G' at low $\omega$. Solid line: Single-mode Maxwell fit ($G_1$, $\tau_1$). Broken line: Sum of two Maxwell modes ($G_1$, $\tau_1$, $G_2$, $\tau_2$). The inset shows the ratio $G_2/G_1$ vs. $\Phi_{si}$.

**Figure 8:** Scattered intensity I as a function of wave vector q for the microemulsion gel (system A) loaded with contrast-matched silica nanoparticles: **A)** $\Omega = 0$; $\Gamma =0.05$, $\Phi_m = 0.04$, r = 10, $\rho_s = 3.5 \ 10^{10}$ cm$^{-2}$. Silica volume fraction $\Phi_{si}= 1$ to 6%. The inset shows a linear representation of the data. **B)** Comparison of sample $\Phi_{si}=6\%$ with corresponding pure silica solution (shifted vertically by a factor of 0.1 for convenience) and pure unconnected swollen micelles at same concentration and rescaled to same contrast.

**Figure 9:** Scattered intensity I as a function of wave vector q for the microemulsion gel (system B) loaded with contrast-matched silica nanoparticles: **A)** $\Omega = 0.4$; $\Gamma$ =0.7, $\Phi_m = 0.20$, r = 10, $\rho_s = 3.5\ 10^{10}\ cm^{-2}$. Silica volume fraction $\Phi_{si}$= 1 to 5%. The inset shows a linear representation of the data. **B)** Comparison of sample $\Phi_{si}$=5% (circles) with corresponding pure silica solution, pure unconnected droplets at same concentration, and silica-free microemulsion gels rescaled to same contrast.

**Figure 10:** Scattered intensity I as a function of wave vector q for the microemulsion gel (system A) loaded with contrast-matched silica nanoparticles at 6%v. The data are compared to two model calculations as explained in the text. A schematic representation of the silica-microemulsion interactions for both cases is also plotted.

**Figure 11:** Schematic view of one elastic cluster made of polymers (curved black lines) and droplets (small striped disks) linked to silica particles (large grey disks). Droplets that are not included in this cluster are not drawn.

**Figures**

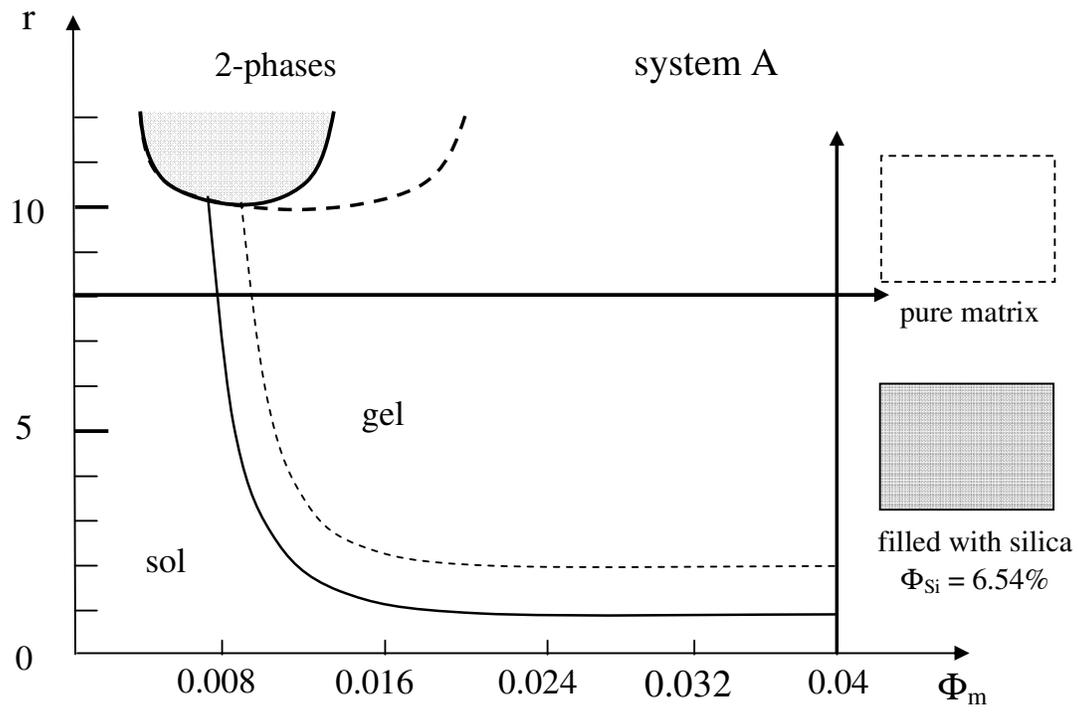

Figure 1 (Puech et al)

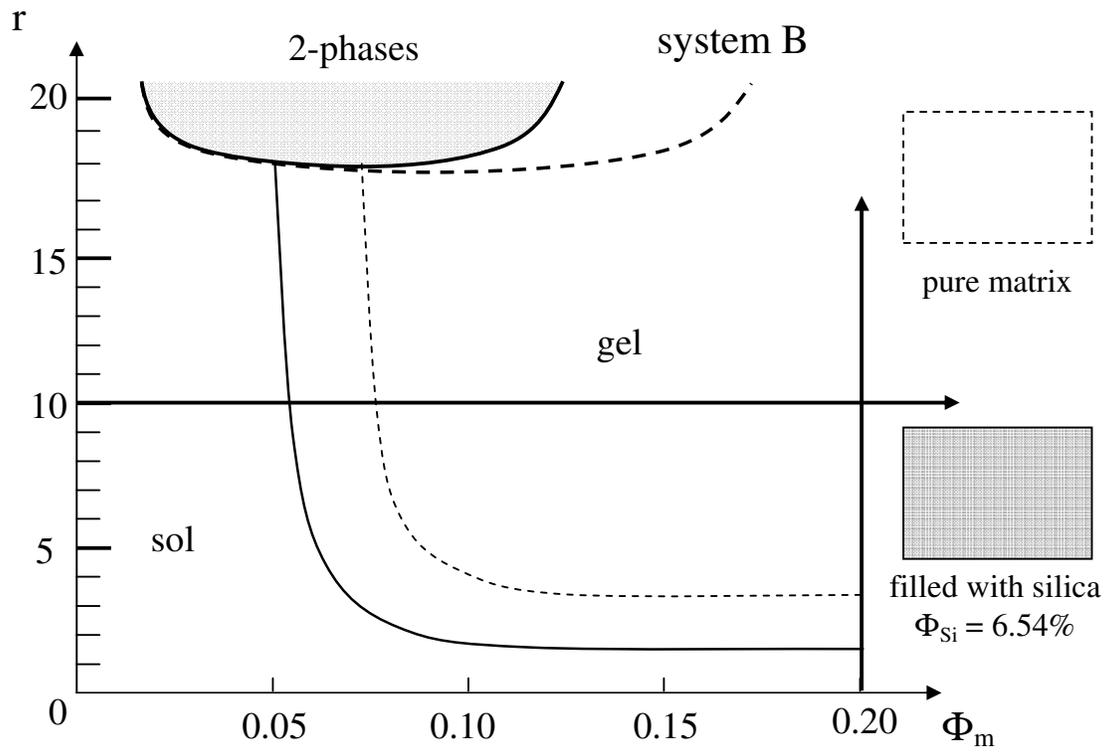

Figure 2 (Puech et al)

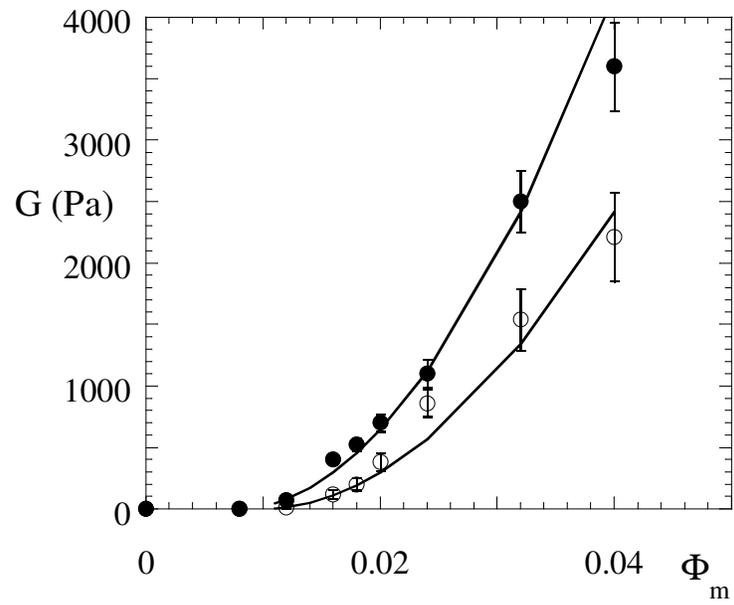

Figure 3 (Puech et al)

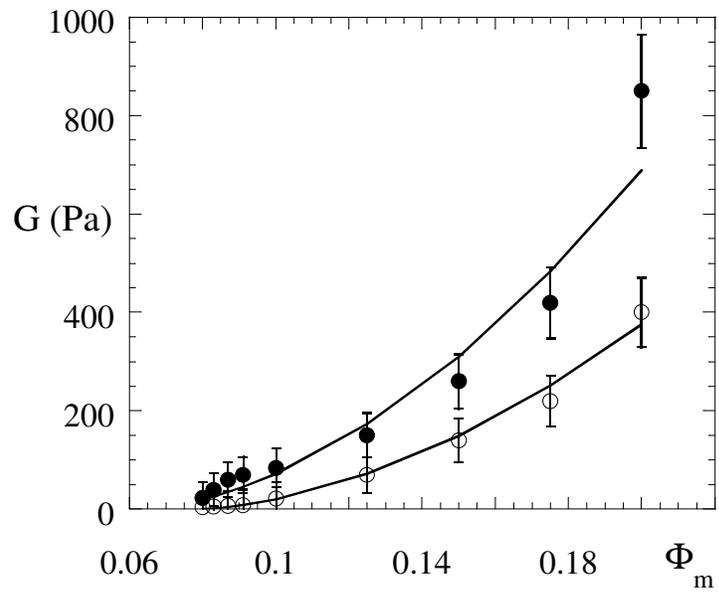

Figure 4 (Puech et al)

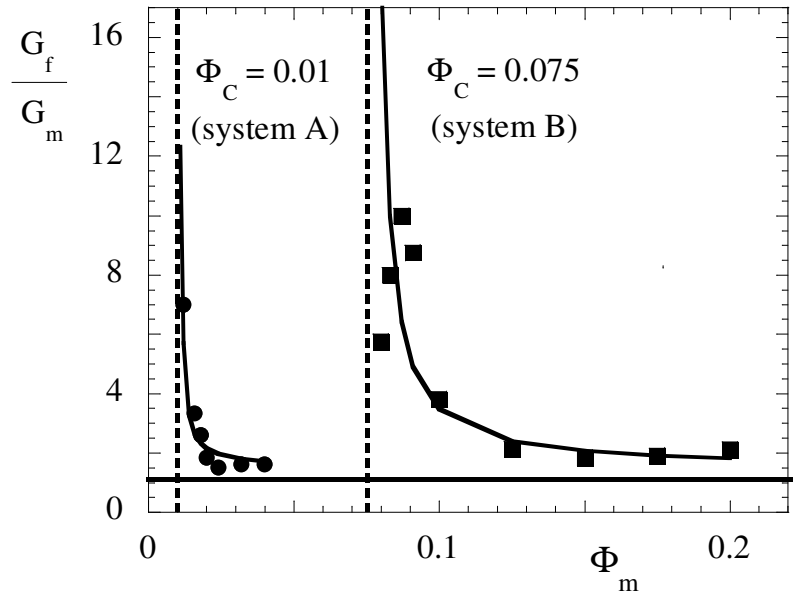

Figure 5 (Puech et al)

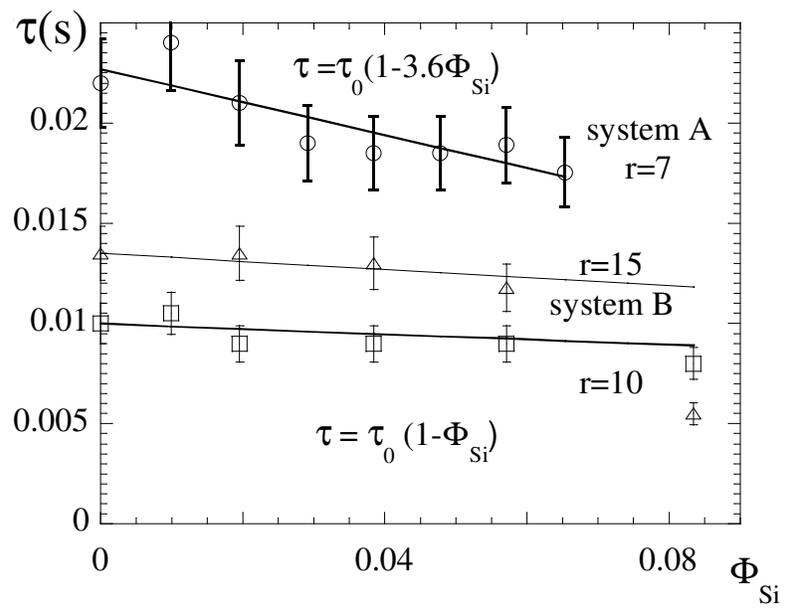

Figure 6 (Puech et al)

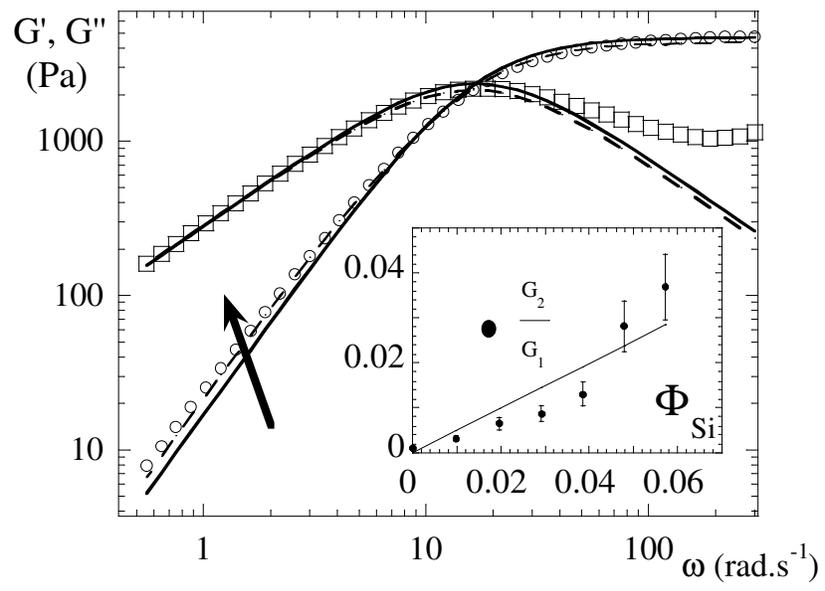

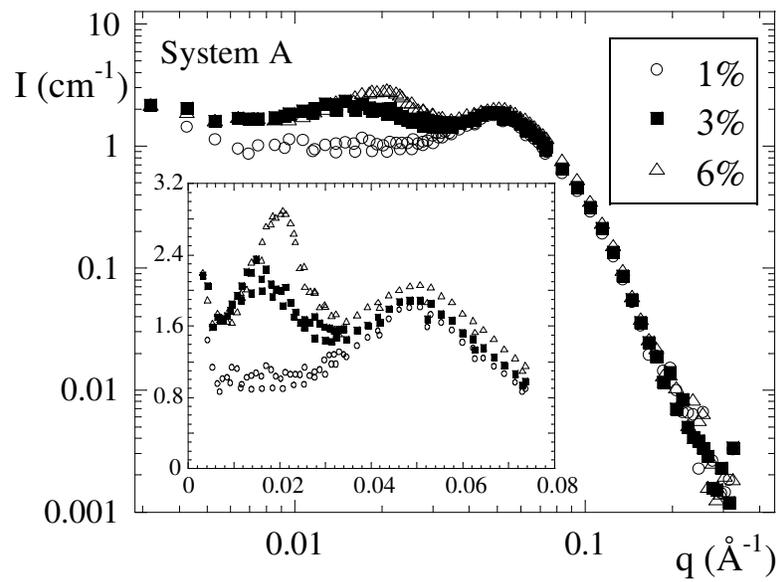

(8a)

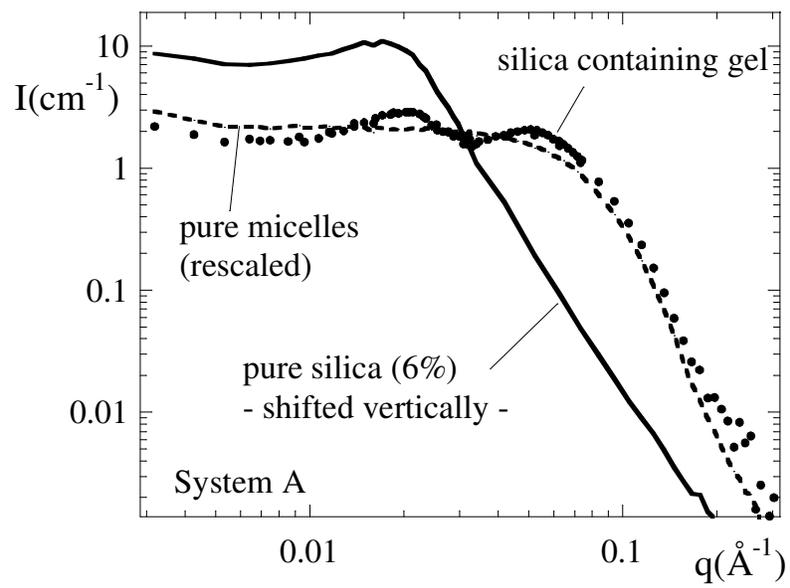

(8b)

Figure 8 (Puech et al)

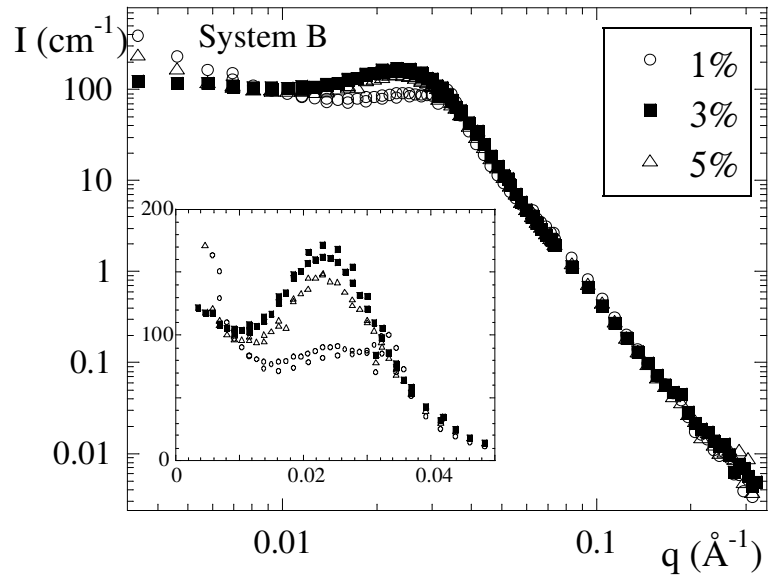

(9a)

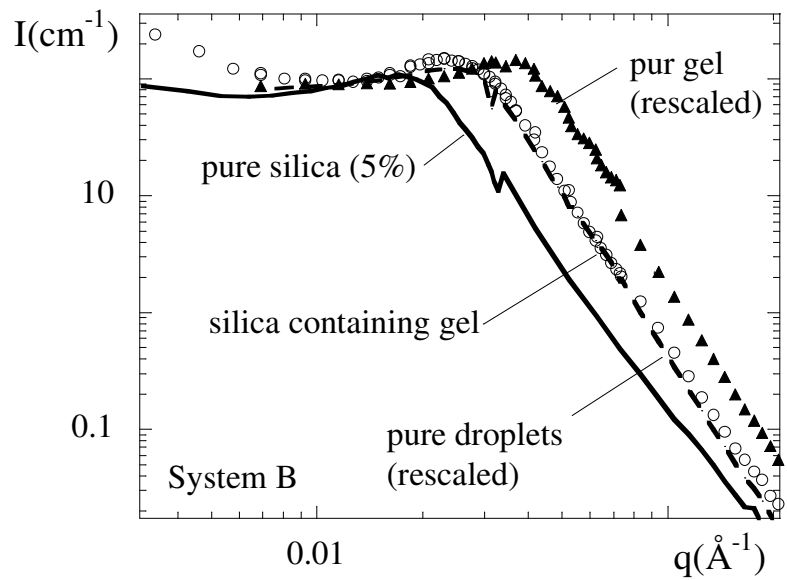

(9b)

Figure 9 (Puech et al)

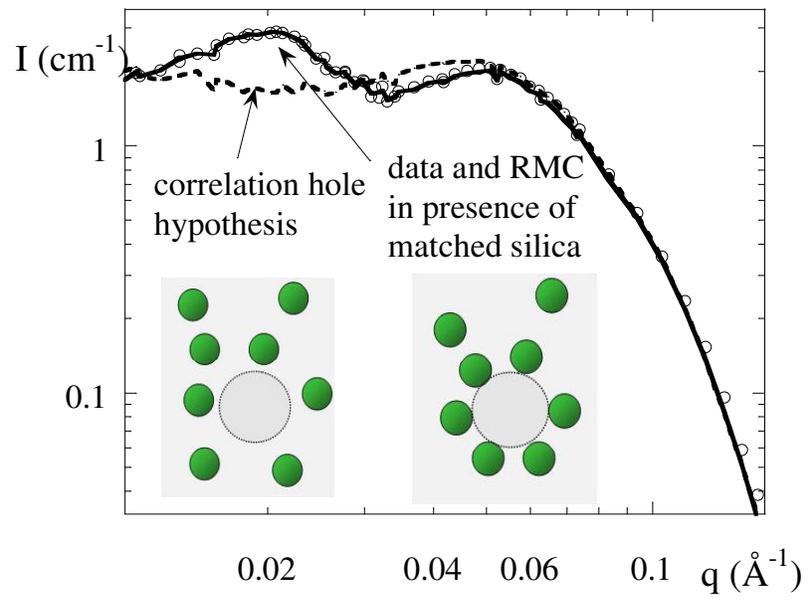

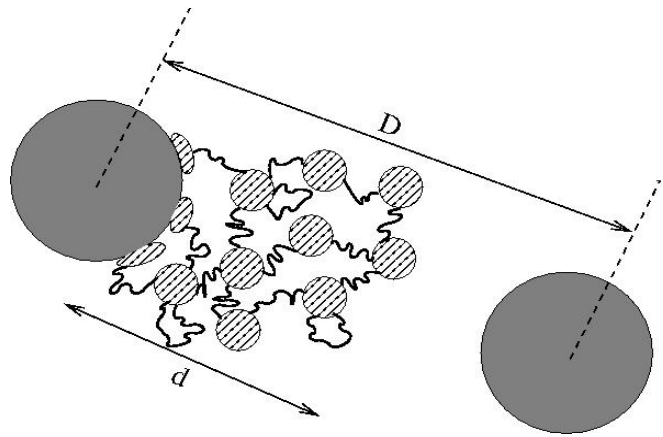

Figure 11 (Puech et al)